\documentclass[conference]{IEEEtran} 

\usepackage{graphicx}
\usepackage{subfigure}
\usepackage{booktabs}
\usepackage{hyperref}
\usepackage{amsmath}
\usepackage{amssymb}
\usepackage{mathtools}
\usepackage{amsthm}
\usepackage{diagbox}
\usepackage{mathtools}
\usepackage{xcolor}
\usepackage{caption}
\usepackage{tabularray}
\usepackage{multirow}

\UseTblrLibrary{booktabs}

\DeclarePairedDelimiter{\ceil}{\lceil}{\rceil}

\theoremstyle{plain}
\newtheorem{theorem}{Theorem}
\newtheorem{proposition}{Proposition}

\theoremstyle{definition}
\newtheorem{definition}{Definition}

\theoremstyle{remark}
\newtheorem{remark}{Remark}

\title{Inference Privacy: Properties and Mechanisms}

\author{\IEEEauthorblockN{Fengwei Tian \; \; Ravi Tandon}
\centering
\IEEEauthorblockA{{Department of Electrical and Computer Engineering} \\
{University of Arizona, Tucson, AZ, 85721}\\
E-mail: \{\textit{fengtian, tandonr}\}@arizona.edu}
}

\begin{document}
\maketitle

\begin{abstract}

    Ensuring privacy during inference stage is crucial to prevent malicious third parties from reconstructing users’ private inputs from outputs of public models. 
    Despite a large body of literature on privacy preserving learning (which ensures privacy of training data), there is no existing systematic framework to ensure the privacy of users’ data during inference. 
    Motivated by this problem, we introduce the notion of Inference Privacy (IP), which can allow a user to interact with a model (for instance, a classifier, or an AI-assisted chat-bot) while providing a rigorous privacy guarantee for the users’ data at inference. 
    We establish fundamental properties of the IP privacy notion and also contrast it with the notion of Local Differential Privacy (LDP). 
    We then present two types of mechanisms for achieving IP: namely, input perturbations and output perturbations which are customizable by the users and can allow them to navigate the trade-off between utility and privacy.
    We also demonstrate the usefulness of our framework via experiments and highlight the resulting trade-offs between utility and privacy during inference.
    
\end{abstract}

\section{Introduction}

    \noindent 
    Machine learning systems, often trained on personalized data and designed to continually interact with users, have become integral to a variety of applications. This opens up a variety of concerns related to user and data privacy. Within machine learning systems, data privacy breaches can occur during both training and inference phases, potentially undermining the system's integrity and performance. While research efforts have primarily focused on understanding and mitigating threats to training data \cite{rigaki2023survey} privacy—such as membership inference and model inversion attacks—addressing privacy challenges during the inference phase remains equally critical.
    In response to such privacy concerns, Differential Privacy (DP) \cite{dwork2014algorithmic} emerged as a standard framework offering a privacy guarantee for training data.
    By introducing controlled noise into the training pipeline, DP mechanisms  can provide a provable guarantee that the trained model will not reveal sensitive information about any specific individual in the training dataset \cite{abadi2016deep}\cite{ghazi2021deep}. 
    
    \noindent While significant attention has been devoted to addressing privacy risks in the training phase, comparatively less focus has been placed on protecting data privacy during inference \cite{jegorova2022survey}\cite{salamatian2015managing}. 
    However, recent studies have shown that a malicious party can reconstruct the input data by observing the model's outputs, posing a privacy threat at inference \cite{malekzadeh2022vicious}.
    Motivated to extend privacy guarantees beyond the training phase to encompass inference phase, we introduce the concept of \textit{\textbf{Inference Privacy }(IP)}. 
    We establish some fundamental properties of IP and compare it to the well-known notion of Local Differential Privacy (LDP) \cite{duchi2013local}. 
    We observe that IP is a generalization of LDP. 
    Additionally, to ensure such privacy guarantees, we propose two inference privacy mechanisms: input perturbation and output perturbation. 
    Specifically, output perturbation method involves introducing controlled noise to the model outputs based on its sensitivity, namely its Global Lipschitz Constant, effectively preventing the disclosure of sensitive information.
    We then present experiments to assess the trade-off between utility and privacy during inference, as well as to evaluate the impact of privacy parameters on utility.
    
    \vspace{2pt}
    
    \noindent \textit{\textbf{Main Contributions:}} 
    The primary objective of this work is to establish a framework for privacy protection during inference, formally introducing the concept of inference privacy.
    The key contributions of this paper are summarized as follows:
    
    \noindent \begin{itemize}
    
        \item \noindent We introduce the concept of Inference Privacy (IP), a new framework designed to ensure privacy for user's query/input data during inference. 
        \noindent The core idea behind IP is to obscure model outputs to the extent that adversaries are unable to discern the specific query input within a defined privacy radius ($\alpha$).
        
        \item We delineate several properties of the IP notion--
        post-processing, composition, and the chaining property.
        Utilizing the Lipschitz continuity inherent in the pre-trained model, we craft multiple output perturbation methods by introducing noise to alter the model's outputs.
        Moreover, we extend our methods to include input perturbation, which can be universally applied across models, enhancing the applicability of our approach.
        
        \item We experimentally test our inference privacy mechanisms on various datasets, comparing the utility of pre-trained models across different IP requirements and methods. 
        Our findings reveal a trade-off: higher inference privacy requirements often come at the expense of reduced utility.
        
    \end{itemize}
    
    \noindent \textit{\textbf{Related Work:}} 
    Previous research provides valuable insights into protecting against information leakage at different stages and levels \cite{rigaki2023survey}.
    Local Differential Privacy (LDP) \cite{kasiviswanathan2011can} ensures privacy during data collection stage by allowing each user to perturb individual data points before their inclusion in the dataset \cite{duchi2013local}, thereby obscuring user contributions and preventing the identification of specific individuals' data.
    Common data-level techniques include blurring query datasets through methods such as data obfuscation \cite{zhang2018privacy} or data sanitation \cite{tambe2016data}, which selectively removing sensitive features \cite{guo2019certified}. 
    However, these approaches may compromise data integrity, and the absence of such features can itself pose a privacy risk. 
    Model-level strategies often include enhancing trained models \cite{papernot2016distillation} or eliminating sensitive data \cite{golatkar2020eternal}.
    
\section{Inference Privacy}

    \begin{figure}        
    
        \includegraphics[width = 0.5 \textwidth]{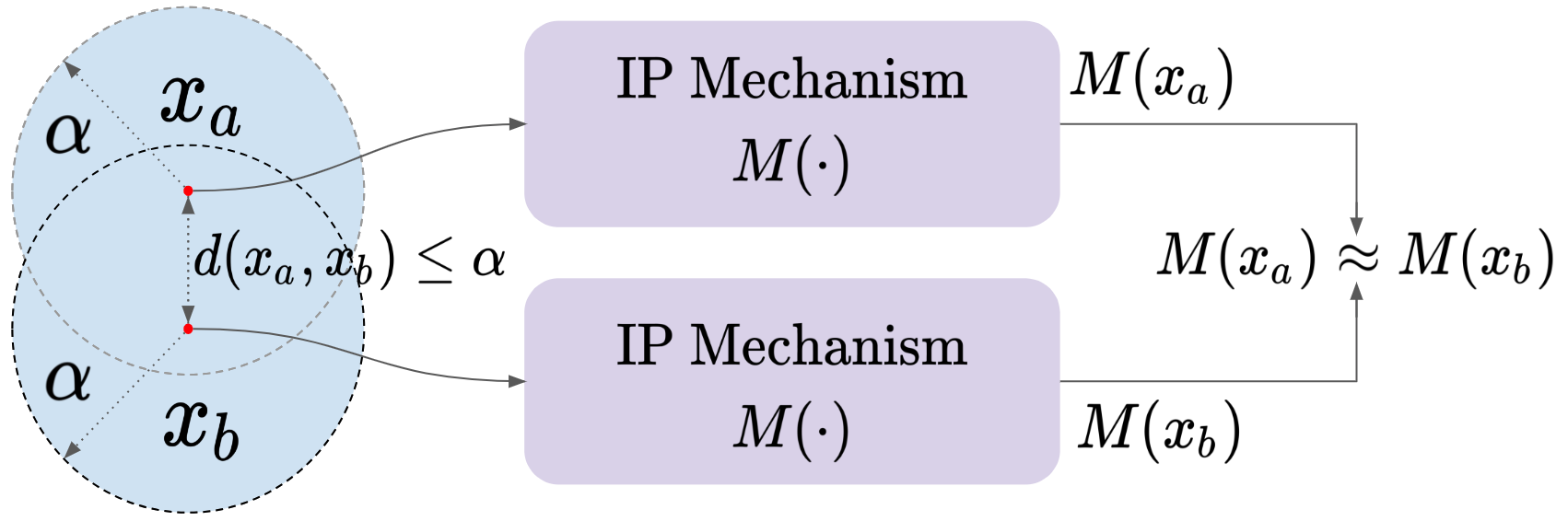}
        \caption{\small{Illustration of Inference Privacy (IP): A mechanism satisfies IP if for any two inputs $x_a$ and $x_b$, such that $||x_a -x_b||_p \leq \alpha$, their corresponding outputs $M(x_a)$ and $M(x_b)$ have similar probability distributions. The radius $\alpha$ measures the extent of similarity, and the privacy leakage is measured by parameters $(\epsilon, \delta)$. (See definition \ref{Approximate}.)}}
        \label{Illustration}
        \vspace{-10pt}
        
    \end{figure}

    \noindent We consider an arbitrary pre-trained model $C: \mathbb{R}^{n} \to \mathbb{R}^{k}$ at inference, where $x \in \mathbb{R}^{n}$ and $C(x) \in \mathbb{R}^k$ signifies the output. 
    The resulting output $C(x)$ can serve multiple downstream applications, such as personalized healthcare recommendations, financial predictions, and other forecasting systems tailored based on user preferences and other sensitive information.
    The goal of IP is to provide a privacy guarantee for the input $x$, so that the reconstruction on $x$ based on $C(x)$ is obfuscated while still preserving high model utility.  
    We now develop the notion of IP as follows: Let us denote $M(x)$ as a randomized mechanism used to privately release $C(x)$ for a private input $x\in \mathcal{X}$, and a metric $d: \mathcal{X} \times \mathcal{X} \rightarrow \mathbb{R}$ that quantifies distance between different inputs. 
    In the $(\mathcal{X}, d)$ metric space, the closed ball centred at $x$ with radius $\alpha$ is defined as:
    \vspace{-2pt}
    \begin{align}
        B(x, \alpha) := \{x'\in \mathcal{X} | d(x', x) \leq \alpha \} .
    \end{align}
    In context of an Euclidean $p$-space, the closed $\ell_p$-ball of radius $\alpha$ centred at $x$, denoted as $B_p(x, \alpha)$ is defined as:
    \begin{align}
        B_p(x, \alpha) := \{x' \in \mathcal{X}| ||x -x'||_p \leq \alpha\} .
    \end{align} 
    
    \noindent Illustrated in Figure \ref{Illustration}, for any two neighboring data inputs $x_a$ and $ x_b$ such that $ d(x_a, x_b) = ||x_a-x_b||_p \leq \alpha$, their corresponding output of IP mechanism $M(x_a)$ and $M(x_b)$ exhibit similar probability distributions. 
    The degree of privacy leakage is associated with output similarity, while the parameter $\alpha$ quantifies the extent of input similarity, offering users the flexibility to adjust it according to the specific input $x$. 
    For instance, if users perceive $x$ as highly private or highly heterogeneous in nature, they might opt for a larger value of $\alpha$ in contrast to a scenario where $x$ is deemed less private or highly homogeneous. 
    The second ingredient is how to measure the similarity between $x_a$ and $x_b$ for all pairs of $x_a, x_b \in B(x, \alpha)$. 
    The choice of metric heavily depends on the specific privacy definition in different contexts.     
    Specifying the metric used to gauge similarities is crucial for providing a meaningful privacy guarantees. 
    Combining these concepts, we then arrive at the following definition of IP.
    
    \begin{definition} [Pure $\{\epsilon, \alpha\}$ Inference Privacy] 
        \label{Pure}
         A randomized mechanism $M$ satisfies $\{\epsilon, \alpha\}$ inference privacy with respect to a metric $d$, if for all measurable sets $S \subseteq Range(M)$:  
        \begin{align}
             Pr[M(x_a)\in S] \leq & \ e^{\epsilon} { Pr[M(x_b)\in S]} ,
        \end{align}
        for all $x_a, x_b$ such that $d(x_a, x_b) \leq \alpha$ .
    \end{definition}
    
    \noindent More generally, we can relax our privacy constraint by introducing a privacy loss parameter $\delta$:
    
    \begin{definition}[Approximate $\{(\epsilon, \delta),\alpha\}$ Inference Privacy]       
        \label{Approximate}
        A randomized mechanism $M$ satisfies $\{(\epsilon, \delta),\alpha\}$ inference privacy with respect to a metric $d$, if for all measurable sets $S \subseteq Range(M)$:
        \begin{align}
             Pr[M(x_a)\in S] \leq & \ e^{\epsilon} { Pr[M(x_b)\in S]}+ \delta  ,
        \end{align}
        for all $x_a, x_b$ such that $d(x_a, x_b) \leq \alpha$ .
    \end{definition}
    
    \noindent We note that pure $\{\epsilon, \alpha\}$ IP is a special case of $\{(\epsilon, 0),\alpha\}$ IP.
    
    \begin{remark} [Comparison with Local Differential Privacy]
        In prior research, the LDP framework has emerged as a standard approach for secure data collection. 
        LDP notion offers a privacy guarantee \cite{kasiviswanathan2011can} that for a randomized algorithm $M$ satisfies $(\epsilon, \delta)$ LDP if for all measurable sets $S \subseteq Range(M)$:
        \begin{align}
            Pr[M(x_a) \in S] \leq e^{\epsilon} Pr[M(x_b) \in S] + \delta ,
        \end{align}
        
        \noindent for all $x_a, x_b$. 
        The $(\epsilon, \delta)$ LDP framework offers robust privacy assurances across all pairs of inputs $x_a$ and $x_b$.
        In contrast, the $\{(\epsilon, \delta), \alpha\}$ IP framework extends the scope of privacy guarantees, where privacy is bounded within the closed $\ell_p$-ball of radius $\alpha$. 
        This expansion broadens the applicability of privacy protection beyond LDP. 
        Notably, $(\epsilon, \delta)$ LDP emerges as a specific instance of $\{(\epsilon, \delta), \infty\}$ IP.
        Thus, $\{(\epsilon, \delta), \alpha\}$ IP is a generalization of $(\epsilon, \delta)$ LDP at inference stage. 
    \end{remark}

    \begin{remark} [Selection of Metric]
        It's important to note that $\{(\epsilon, \delta), \alpha\}$ IP only ensures privacy with respect to a chosen metric $d$, and the selection of this metric is pivotal in determining the radius $\alpha$. 
        The selection of a metric for addressing privacy concerns is contingent upon the inherent characteristics of the input data. 
        For instance, in text-based classification scenarios where semantic interpretation is paramount, opting for a semantic similarity metric between two word vector representations is common \cite{church2017word2vec}, contrasting with the Levenshtein distance, which measures the dissimilarity between two words based on single-character edits \cite{yujian2007normalized}. 
        For example, while the words "uninformed" and "uneducated" have similar semantic meanings, the Levenshtein distance between them are quite large, but the Levenshtein distance between "uninformed" and "uniformed" is merely 1 while they have very distinctive semantic meanings.
        However, it's crucial to recognize that these distances inherently rely on discrete representations owing to the discrete nature of language.
        In contrast, when dealing with images, visual features take precedence, leading to the adoption of metrics like Euclidean distance or Structural Similarity Index (SSIM). 
        Here, the choice of metric is closely tied to the specific task at hand, with an emphasis on capturing visual similarity for image-related tasks. 
    \end{remark}

    \begin{remark} [Comparison with Metric Differential Privacy]
        The concept of providing privacy guarantees based on the radius $\alpha$ between pairs of inputs also has connections to Metric Differential Privacy ($d_{\mathcal{X}}$-DP) \cite{alvim2018local} and its recent extension at the user level \cite{imola2024metric}. 
        Informally, a mechanism $M$ satisfies bounded $d_{\mathcal{X}}$-DP if, for all $x_a, x_b$ and for all measurable sets $S \subseteq Range(M)$:
        \begin{align}
            Pr[M(x_a) \in S] \leq e^{\epsilon d (x_a, x_b)} Pr[M(x_b) \in S] + \delta . 
        \end{align}
        This definition generalizes LDP when  $ d (x_a, x_b) = 1$, offering privacy guarantee based on the selected metric and the distance between two inputs.  
        We consider the IP framework  a more robust generalization of $d_{\mathcal{X}}$-DP, as IP framework addresses the impact on privacy guarantees of the distance between two ``adjacent" data entries.
        Moreover, our focus is on ensuring data privacy during the inference stage rather than the data collection stage, distinguishing our work from prior research.
    \end{remark}

        

\section{Properties of Inference Privacy} 
    
    \noindent The mathematical definition of IP bears resemblance to the definition of Differential Privacy (DP). 
    Consequently, it is pertinent to explore whether IP exhibits similar properties to those of DP.
    IP demonstrates resilience to \textbf{post-processing} meaning that any attempt by a data analyst to derive a function from the output $M(x)$ of an IP mechanism $M$ with the intent of reducing its privacy guarantee is fruitless unless additional knowledge about the private input $x$ is available. 
    In essence, if an algorithm ensures IP at the initial stage of an inference process, the same level of privacy guarantee is upheld throughout the entirety of the inference process.
    Formally, we show that the composition of a data-independent mapping function $F$ with an $\{(\epsilon, \delta), \alpha\}$ inference private mechanism $M$ is also $\{(\epsilon, \delta), \alpha\}$ inference private. 
    
    \begin{proposition}[Post-Processing Property of IP Mechanisms]
        Let $M: \mathbb{R}^n \rightarrow \mathbb{R}^k$ be a randomized algorithm that satisfies $\{(\epsilon, \delta), \alpha\}$ IP. 
        Let $F:\mathbb{R}^k \rightarrow \mathbb{R}^{k'}$ be an arbitrary randomized mapping.
        Then, $F \circ M :\mathbb{R}^n \rightarrow \mathbb{R}^{k'}$ satisfies  $\{(\epsilon, \delta), \alpha\}$ IP.
    \end{proposition} 
    
    \noindent Proof of this result is presented in Appendix \ref{Post Processing}. 

    \begin{remark} [Limitation of Post-Processing Property]
        The post-processing property of IP permits the application of additional computations or transformations to the output of IP algorithms without compromising the established privacy guarantees of the original computation. 
        However, relying solely on post-processing may not suffice to ensure privacy in real world applications. 
        While the post-processing property offers privacy assurance for individual tasks, it overlooks the potential privacy loss spring from the sequential performance of multiple independent computations on the same data input. 
        In contrast, the composition of independent mechanisms is a facet of IP that tackles this concern by quantifying the aggregate privacy loss.
        Formally, we show that the \textbf{basic composition} of multiple independent IP mechanisms is also inference private. 
    \end{remark}

    \begin{proposition}[Basic Composition of Independent IP Mechanisms]
        Let $M_i : \mathbb{R}^{n} \rightarrow \mathbb{R}^{k}$ be an $\{(\epsilon_i, \delta_i), \alpha_i\}$ inference private algorithm. 
        Then if $M$ is defined to be $M(x) = (M_1(x), ..., M_m(x))$, where each $M_i$ are independent from each other. 
        Then $M$ satisfies $\{ ( \sum_{i=1}^m \epsilon_i,  \sum_{i=1}^m \delta_i, \min \alpha_i \}$ IP. 
    \end{proposition}
    \noindent Proof of this result is presented in Appendix \ref{Basic Composition}. 
    
    \begin{remark} [Limitation of Basic Composition]
        The basic composition of independent mechanisms ensures overall privacy guarantees when multiple IP computations occur sequentially on the same data input. 
        This enables the division of the inference stage into distinct tasks, allowing for a balance between total privacy guarantee and task utility. 
        However, it doesn't address privacy loss when different computations occur simultaneously on different data subdivisions. 
        In contrast, \textbf{parallel composition} tackles this concern by splitting the data input into distinct partitions. 
        Formally, we show that the parallel composition of multiple independent IP mechanisms is also inference private.
    \end{remark}
    
    \begin{proposition}[Parallel Composition of Independent IP Mechanisms]
        Let an arbitrary input $x \in \mathbb{R}^n$ be spited into $m$ disjoint chunks such that $x = x_1 \cup x_2 ... \cup x_m $ and $x_i \cap x_j = \emptyset$ for any $x_i \neq x_j$ , where each $x_i \in \mathbb{R}^{n_i}$ and $\sum^m_{i=1} n_i = n$. 
        Let $M_i : \mathbb{R}^{n_i} \rightarrow \mathbb{R}^{k_i}$ be an $\{(\epsilon_i, \delta_i), \alpha_i\}$ inference private algorithm for each partition $x_i \in \mathbb{R}^{n_i}$. 
        Then if $M$ is defined to be $M(x) = (M_1(x_1), ..., M_m(x_m))$, where each $M_i$ are independent from each other. 
        Then $M$ satisfies $\{ ( \sum_{i=1}^m \epsilon_i,  \sum_{i=1}^m \delta_i), \min \alpha_i \}$ IP. 
    \end{proposition}
    \noindent Proof of this result is presented in Appendix \ref{Parallel Composition}. 
    
    \begin{remark}[Comparison with DP Parallel Composition]
        In the previously proposed DP framework, the parallel composition theorem ensures that when applying $m$ DP mechanisms $M_1$, $M_2, ..., M_m$ computed on disjoint subsets of the private database, each mechanism provides differential privacy guarantees of $(\epsilon_1, \delta_1)$,  $(\epsilon_2, \delta_2)$, ..., $(\epsilon_m, \delta_m)$ respectively, then the composed mechanism $M$ defined as $M(x) = (M_1(x_1), ..., M_m(x_m))$, adheres to $( \max_{i=1}^m \epsilon_i,  \max_{i=1}^m \delta_i)$ DP. 
        In contrast with the DP framework, the parallel composition of IP operates differently.
        Differential privacy primarily concerns datasets that differ by only one element, resulting in nearly identical partitions except for one subset. 
        The privacy guarantee of the composed mechanism relies solely on this differing subset.
        Conversely, in the parallel composition of IP framework, all partitions deviate from their counterparts. 
        Consequently, the privacy guarantee of the composed mechanism is influenced by all partitions, leading to a higher privacy budget compared to individual subdivisions.
    \end{remark}
    
    \begin{remark} [Application of Parallel Composition]
       The parallel composition property of IP presents an opportunity to devise strategies that strategically apply distinct privacy requirements to various subdivisions of an input, all while maintaining the total privacy budget unaltered.
       Consider a scenario involving text-based data input. 
       Here, we can implement a more stringent privacy constraint on sensitive keywords by leveraging both arbitrary selection and focused attention on the texts. 
       Simultaneously, we can afford to relax the privacy requirement on less crucial text segments. 
       This deliberate adjustment allows for an enhancement in the utility of the model while ensuring that the overall privacy budget remains intact.
       This capacity to selectively tailor privacy provisions to different components of the input opens avenues for privacy management, enabling organizations to balance privacy concerns with the optimization of utility across diverse data domains and applications.
    \end{remark}

    \noindent Parallel composition offers a privacy guarantee for a partitioned data input, with each subdivision potentially having a distinct radius $\alpha_i$.  
    It's valuable to investigate how changes in the radius affect the privacy guarantee.
    The \textbf{chaining property} of the IP framework allows for the extension of the radius under the same IP mechanism. 
    This attribute enables the application of an IP mechanism across various scenarios, thereby broadening the scope of privacy protection beyond individual subdivisions.
    Formally, we define the chaining property of inference private mechanism as:
    \begin{proposition}[Chaining Property of IP mechanism in Euclidean $p$-Space]
        Let $M : \mathbb{R}^{n} \rightarrow \mathbb{R}^{k}$ be an $\{(\epsilon, \delta), \alpha\}$ inference private algorithm. 
        Then $M$ satisfies $\{(\ceil{\frac{\beta}{\alpha}}\epsilon,  [\frac{e^{\ceil{\frac{\beta}{\alpha}}\epsilon}-1}{e^{\epsilon}-1} ] \delta), \beta \}$ IP for any $\beta \geq 0$.
    \end{proposition}
    \noindent Proof of this result is presented in Appendix \ref{Chaining Property}. 
    
    \begin{remark} [Application of Chaining Property]
        The chaining property of IP highlights the balance between the desired privacy radius $\alpha$, and the strength of the privacy guarantee $(\epsilon, \delta)$ within that radius. 
        Mechanism $M$ offers a spectrum of distinct IP guarantees within the corresponding contours.
        Generally, as the interested radius $\alpha$ expands, the privacy budget parameter $\epsilon$ increases linearly, while the privacy loss parameter $\delta$ escalates exponentially.
        This property of IP facilitates a flexible selection of appropriate privacy parameters. 
        Users can tailor these parameters based on specific circumstances. 
        For instance, in scenarios where the data input is predominantly homogeneous, a stronger privacy guarantee is preferable at the expense of a smaller interest radius. 
        Conversely, in situations characterized by highly heterogeneous data, a larger interest radius is favored.
        Moreover, the chaining property of IP resembles many similarity with group differential privacy, which focuses on two datasets differing $k$ elements \cite{dwork2014algorithmic}.  
        The chaining property of IP ensures that the same mechanism remains applicable across diverse scenarios, offering flexibility in mechanism design and accommodating varying privacy needs.
    \end{remark} 
    
\section{Mechanisms for Inference Privacy}

    \begin{figure}
        \includegraphics[width = 0.5 \textwidth]{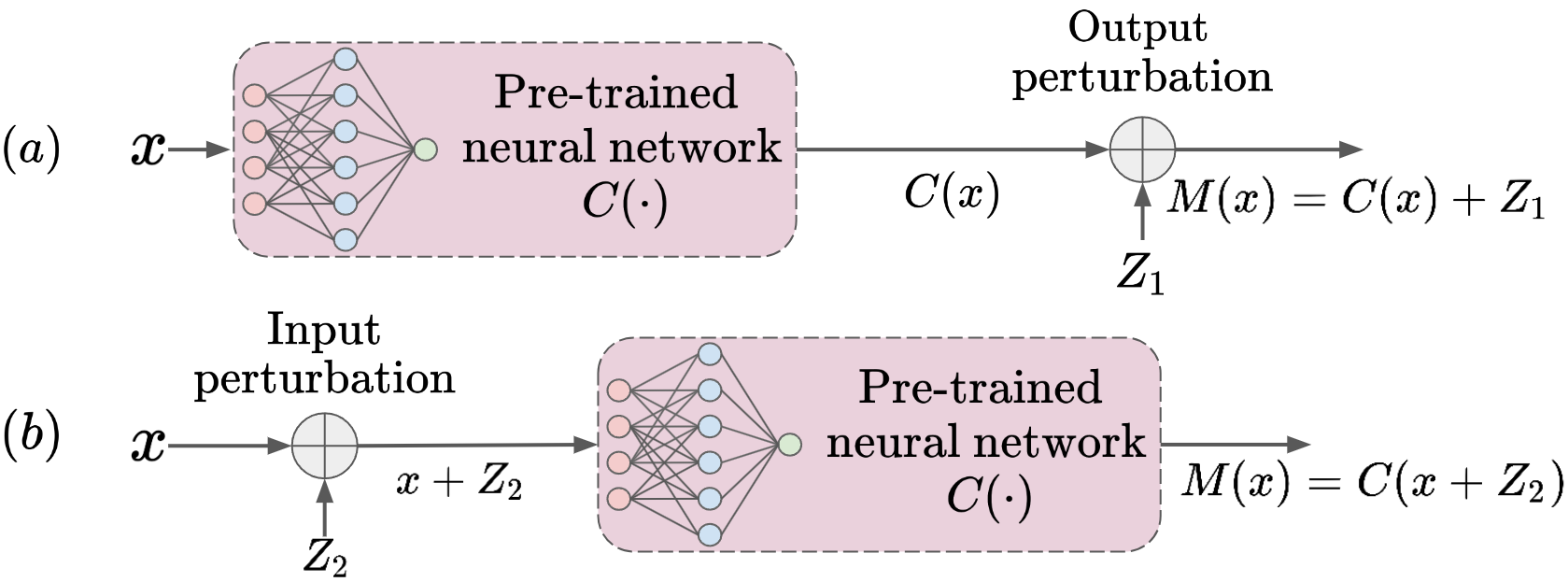}
        \caption{The workflow of output perturbation methods: User generates a noise $N$ based on the model used for inference, and perturbs the model output before releasing it. }
        \label{Methods}
        \vspace{-10pt}
    \end{figure}
    
    \noindent In this section, we present two main approaches of designing IP mechanisms, output perturbation and input perturbation, as shown in Figure \ref{Methods}. 
    
    \noindent \textbf{\textit{Output Perturbation}} is based on injecting controlled noise into the model’s output.
    As illustrated in Figure \ref{Methods}, we use an additive noise $Z_1 \in \mathbb{R}^k$ to corrupt the models output $C(x)$ :
    \begin{align}
        M(x) = C(x) + Z_1,
    \end{align}
    where $Z_1$ is a $k$-dimensional vector with i.i.d. entries drawn from the same probability distribution. 
    Corrupting the output of the model makes it mathematically challenging for any potential adversary to infer the private data. 
    The level of noise injected is not only contingent upon the required privacy parameters $\epsilon$, $\delta$, $\alpha$, but also hinges on the model sensitivity, measured by the model's \textbf{Lipschitz constant}. 
    A less sensitive model necessitates a larger amount of noise compared to a highly sensitive model under the same privacy requirement. 
    The global Lipschitz constant is a measure of the maximum ratio between the variations of the outputs compared to the variations of the inputs. 
    However, it is computationally infeasible to accurately estimate Lipschitz constants, especially for larger networks. 
    Consequently, it is practical to use upper bounds to approximate these constants.
    For a pre-trained model $C :\mathbb {R}^d \rightarrow \mathbb{R}^k$, the  $\ell_p$ global Lipschitz constant is defined as follows: 
    \vspace{-5pt}
    \begin{align}
        \mu_C = \sup_{x \neq x' }\frac{|| C(x)-C(x')||_p}{||x - x'||_p}  , 
    \end{align}
    For simplicity, we refer to the upper bound of the global Lipschitz constant as the “global Lipschitz constant”, also denoted by $\mu_C$. 
    
    \noindent \textbf{1-Lipschitz Neural Networks} \cite{virmaux2018lipschitz} have been introduced in recent literature as a means to improve the stability of neural networks \cite{fazlyab2019efficient}.
    Through some unified semi-definite programming approach \cite{araujo2023unified}, such as Cayley Transform \cite{trockman2021orthogonalizing} and other orthogonal parameterization approaches\cite{xu2022lot}, these networks aim to ensure the neural network has a global Lipschitz of 1. 
    By guarantee that minor alterations in the input space translate to only minimal changes in the output space, this characteristic promotes smoother and more consistent model behavior, contributing to improved stability overall.
    Output perturbation methods effectively exploit this property, as they rely on the Lipschitz constant to determine the extent of noise injected into the output. 
    For a model with a smaller Lipschitz constant, the injected noise is proportionally smaller, ensuring that the perturbed output remains interpretable and meaningful. 
    However, Lipschitz constants are subject to the $\ell_p$ norm, denoted as $||\cdot||_p$, and the selection of the suitable norm depends on the particular properties and requirements of the problem under consideration. 
    
    \noindent For instance, $\ell_1$ Lipschitz constants are less affected by outliers compared to other norms and are computationally less expensive to calculate.
    In situations where data might include outliers or noise, employing an $\ell_1$ Lipschitz constant can result in more resilient solutions. 
    The \textbf{Lap-Output Mechanism} is suitable to provide an IP guarantee utilizing $\ell_1$ Lipschitz constants. 
    
    \begin{definition}[Lap-Output Mechanism for $\{ (\epsilon, 0), \alpha \}$ IP]
        
    Given any arbitrary pre-trained function $C(\cdot)$ that takes input $x$ and outputs $C(x) \in \mathbb{R}^k $, the Lap-Output Mechanism is defined as:
        \vspace{-10pt}
        \begin{align}
            M_{Lo}(x) = C(x) + Z_1 ,
        \end{align}
        where $Z_1$ is a $k$-dimensional vector with i.i.d. entries drawn from $ Lap(\frac{\mu_C \alpha}{\epsilon})$ and $\mu_C$ is the $\ell_1$ global Lipschitz constant. 
    \end{definition} 
    \noindent We show that the Lap-Output Mechanism satisfies $\{ (\epsilon, 0), \alpha \}$ IP. 
    Poof of this result is presented in Appendix \ref{Lap-Output}. 
    
    \begin{theorem}
        Lap-Output mechanism satisfies $\{ (\epsilon, 0), \alpha \}$ IP.
    \end{theorem}
    
    \noindent Despite its numerous advantages, $\ell_1$ global Lipschitz constant can sometimes be quite large, necessitating a higher level of noise distortion to meet  
    the $\{ (\epsilon, 0), \alpha \}$ IP requirements. 
    To mitigate the distortion and maintain a higher utility level for $C(x)$, the $\ell_2$ Lipschitz constant is often employed. 
    
    \noindent Given the that $\ell_2$ global Lipschitz constant is always smaller or equal to the $\ell_1$ global Lipschitz constant, we anticipate experiencing less distortion in $M(x)$. 
    The \textbf{Gauss-Output Mechanism} is suitable to provide an IP guarantee utilizing $\ell_2$ global Lipschitz constants. 
    However, we must relax IP requirement by incorporating a privacy loss parameter $\delta$. 
    
    \begin{definition}[Gauss-Output Mechanism for $\{ (\epsilon, \delta), \alpha \}$ IP]
        Given any function $C(\cdot)$ takes input $x \in \mathbb{R}^n$ and outputs $C(x) \in \mathbb{R}^k $, the Gauss-Output Mechanism is defined as: 
        \begin{align}
            M_{Go}(x) = C(x) + Z_2, 
        \end{align}
        where $Z_2$ is a $k$-dimensional vector with i.i.d. entries drawn from $ \mathcal{N}(0, \sigma^2)$ ,  $\sigma^2 = 2ln(\frac{1.25}{\delta})\frac{(\alpha {\mu_C)} ^2}{\epsilon^2 }$ and $\mu_C$ is the $l_2$  global Lipschitz constant of $C(x)$.
    \end{definition} 

    \noindent We show the Gauss-Output Mechanism satisfies $\{ (\epsilon, \delta), \alpha \}$ IP. Proof of this result is presented in Appendix \ref{Gauss-Output}.  
    
    \begin{theorem}
       Gauss-Output mechanism satisfies $\{ (\epsilon, \delta), \alpha \}$ IP.
    \end{theorem}
    
    \begin{remark} [Limitation of Output Perturbation Approach]
        The output perturbation approach relies on the model's global Lipschitz constant, which may not always be small or even finite. 
        Additionally, the Lipschitz constant is an intrinsic property that varies for each model. 
        Consequently, all output perturbation mechanisms must be custom-designed to suit each model individually. 
        This lack of universality makes it challenging to devise a one-size-fits-all method applicable in all situations. 
    \end{remark}
    
    \noindent \textbf{\textit{Input Perturbation}} is based on injecting controlled noise into the model's input as illustrated in Figure \ref{Methods}. 
    In contrast to model-tailored output perturbation methods, input perturbation techniques do not depend on the specific model at inference.
    Serving as an universal approach, input perturbation methods offer a simpler mechanism, albeit potentially suffering from utility loss. 
    By introducing additive noise $Z_2 \in \mathbb{R}^n$ to corrupt the model input, we achieve:
    \begin{align}
        M(x) = C(x + Z_2), 
    \end{align}
    where $Z_2$ is a $n$-dimensional vector with i.i.d. entries drawn from the same probability distribution. 
    Corrupting the input of the model renders it mathematically challenging for any potential adversary to deduce the private data. 
    Intuitively, this outcome can be achieved by extending the post-processing property of IP.
    Consider a special arbitrary function $A(\cdot) : \mathbb {R}^n \rightarrow \mathbb {R}^n$ that outputs the models input directly, as well as an output perturbation IP mechanism $M(\cdot)$ such that:
    \begin{align}
        M(x) = A(x) + Z_2 = x + Z_2. 
    \end{align}
    Following the post-processing property, for another arbitrary model $C(\cdot)$, $C \circ M$ must satisfy the same IP requirement. 
    Notice that the parameters of the injected noised are only depended on the IP requirement and the Lipschitz constant for $A(\cdot)$, which is always 1 regardless of $x$. 
    These parameters are invariant to the model $C(\cdot)$, making input perturbation methods universally adaptable. 
    In comparison with the output perturbation approach, the \textbf{Gauss-Input Mechanism} can also provide an IP guarantee. 
    
    \begin{definition}[Gauss-Input Mechanism for $(\{ \epsilon, \alpha \}, \delta)$ IP]
        
       Given any function $C(\cdot)$ takes input $x \in \mathbb{R}^n$ and outputs $C(x) \in \mathbb{R}^k $, the Gauss-Input Mechanism is defined as:
        \begin{align}
            M (x) = C(x + Z_2),
        \end{align}
        where $Z_2$ is a $n$-dimensional vector with i.i.d. entries drawn from $ \mathcal{N}(0, \sigma^2)$, and $\sigma^2 = 2ln(\frac{1.25}{\delta})\frac{\alpha^2}{\epsilon^2 }$. 
    
    \end{definition} 
    
    \noindent We show that the Gauss-Input Mechanism satisfies $\{ (\epsilon, \delta), \alpha \}$ IP.
    Proof of this result is presented in Appendix \ref{Gauss-Input}. 
    
    \begin{theorem}    
        Gauss-Input mechanism satisfies $\{ (\epsilon, \delta), \alpha \}$ IP.
    \end{theorem}

    \begin{remark}[Comparison Between Input Perturbation and Output Perturbation]
        The input perturbation method demonstrates universal adaptability since the injected noise is solely determined by the IP requirements. 
        In contrast, the output perturbation method is customized specifically for the model $C(\cdot)$, as the injected noise depends on both the IP requirements and the global Lipschitz constants $\mu_C$ of the model. 
        However, one can conceptualize input perturbation as a particular instance of output perturbation applied to the private input variable, $x$, in accordance with the post-processing property. 
        It is anticipated to yield comparatively lower utility than output perturbation methods. 
        Nevertheless, given that the model's Lipschitz constant may be notably large, this has spurred research into the development of 1-Lipschitz neural networks.
    \end{remark}

\section{Evaluation}

    \noindent In this section, we present an experimental evaluation of our proposed IP framework focusing on two key questions:
    
    \noindent \textbf{Q1: Impact of various IP mechanisms on model utility:} 
    We investigate how the choice of different IP mechanisms within the IP framework affects the overall utility of the model. 
    Utility, in this context, refers to the model's ability to perform the image classification task accurately.

    \noindent \textbf{Q2: Influence of changing IP parameters on model utility:}
    We analyze how adjustments of various parameters within the IP framework influence the model's utility. 
    This analysis will help us understand the trade-offs between privacy protection and model performance.

    \noindent To address these questions, we conduct a series of experiments where we evaluate different IP mechanisms under varying levels of IP constraints in standard image classification tasks.
    Our code is available at https://github.com/FTian-UArizona/Inference\_Privacy.

    \noindent \textbf{Experiment Settings.}
    We evaluate Inference Privacy framework on image classification of two standard datasets: CIFAR-10 \cite{krizhevsky2009learning} and CIFAR-100 \cite{krizhevsky2009learning}.
    We report the standard classification accuracy for both datasets as the metric for model utility. 
    We utilize several pre-existing ResNet-18 architecture models \cite{rw2019timm} as well as pre-existing \textit{SDP-based Lipschitz Layer (SLL)} networks proposed recently \cite{araujo2023unified} \cite{wang2023direct}. 
    We employed Gauss-Input mechanism on ResNet-18 models and both Gauss-Input mechanism and Gauss-Output mechanism on SLL models. 

    \noindent \textbf{``Optimally" tuned ResNet-18 model.} 
    In the Gauss-Input mechanism, input data is perturbed with noise before being fed into the model to enhance privacy.
    However, pre-trained models are typically not optimized for handling such noisy inputs.
    Fine-tuning the model with noisy data can potentially improve its utility and performance.
    Since the privacy requirement at inference time is usually known, fine-tuning can be specifically targeted to these noise levels.
    To achieve this, we conduct a grid search by fine-tuning the pre-trained model with images subjected to varying privacy levels to identify the bet fine-tuned model with highest accuracy when validated at targeted privacy requirement.

    \begin{figure}
        \centering
        \includegraphics[width = 0.48 \textwidth]{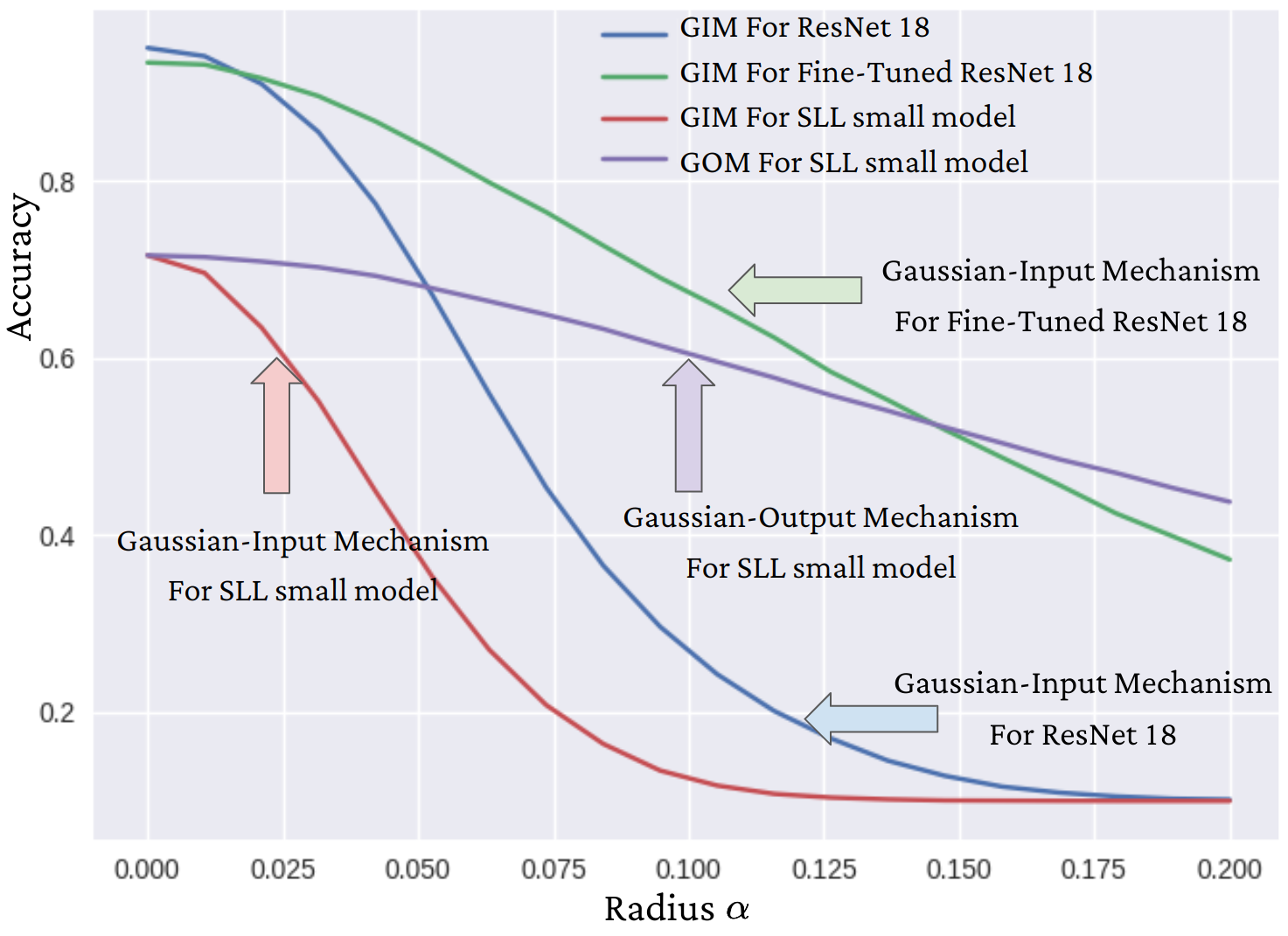}
        \caption{Experimental results on CIFAR-10 classification: natural classification accuracy for input and output perturbation methods as a function of radius $\alpha$ for a fixed $\epsilon = 1$ and a fixed $\delta = 10^{-5}$. Values reported are average of 15 tests. }
        \label{Accuracy_vs_Radius_CIFAR10}
        \vspace{-10pt}
    \end{figure}

    \noindent \textbf{Impact of varying radius $\alpha$ for a fixed privacy budget $\epsilon$.} \\
    As shown in Figure \ref{Accuracy_vs_Radius_CIFAR10}, we evaluated the trade-off between model utility and privacy, demonstrating that as radius $\alpha$ increases from 0 to 0.2, the model classification accuracy decreases for a fixed $\epsilon = 1$ and $\delta = 10^{-5}$.  
    Notably, the Gauss-Output mechanism demonstrates superior performance compared to the Gauss-Input mechanism in the SLL model. 
    Furthermore, the fine-tuned ResNet-18 model consistently outperforms the original ResNet-18 model across all evaluations.
    An intriguing observation is that the Gauss-Output mechanism applied to the SLL model outperforms the Gauss-Input mechanism when implemented on fine-tuned ResNet-18 models in the high privacy region, as the radius $\alpha$ increases. 
    However, it is important to note that in the low privacy region, the SLL model exhibits comparatively lower performance than the ResNet-18 model.
    The observed reduction in utility associated with output perturbation methods highlights a potentially valuable direction for further research into the development of 1-Lipschitz models. 
    Nonetheless, Gauss-Output mechanism is tailored for the SLL model, Gauss-Input mechanism offers practicality advantages as it can be applied to other models. 
    In experiments with a pre-trained ResNet-18 model, implementing the Gauss-Output mechanism was challenging due to the model's large global Lipschitz constant. 
    This discovery may inspire increased focus on small Lipschitz models and encourage further research into more stable models.

    \begin{figure}
        \centering
        \includegraphics[width = 0.48 \textwidth]{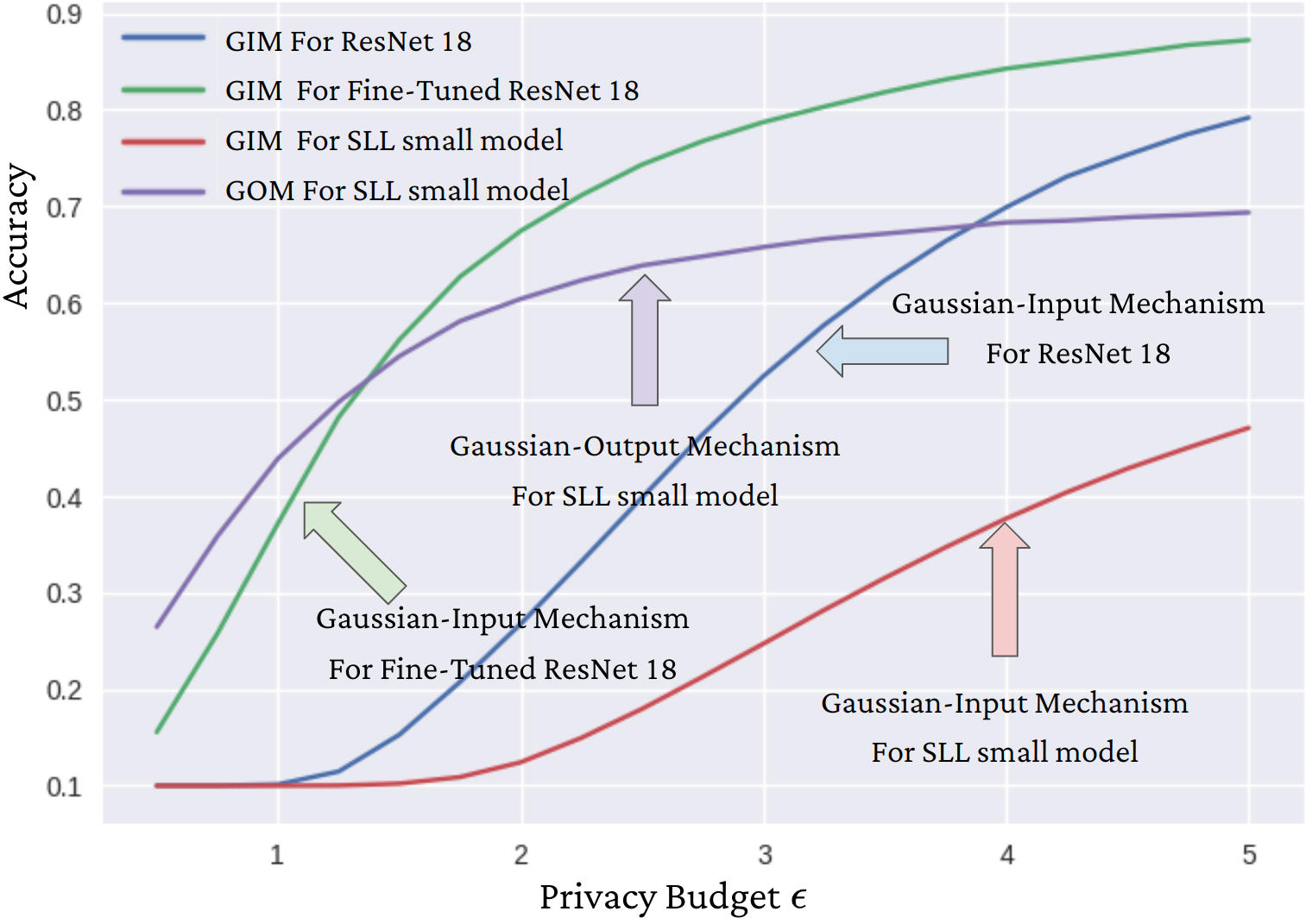}
        \caption{Experimental results on CIFAR-10 classification: natural classification accuracy for input and output perturbation methods as a function of radius $\epsilon$ for a fixed $\alpha = 0.1$ and a fixed $\delta = 10^{-5}$. Values reported are average of 15 tests. }
        \label{Accuracy_vs_Budget_CIFAR10}
        \vspace{-10pt}
    \end{figure}
    
    \noindent \textbf{Impact of varying privacy budget $\epsilon$ for a fixed radius $\alpha$.} \\
    Shown in Figure \ref{Accuracy_vs_Budget_CIFAR10}, we evaluate the trade-off between model utility and privacy budget, demonstrating that as privacy budget $\epsilon$ increases from $0.25$ to $5.0$, the natural classification accuracy decreases for a fixed $\delta$ of $10^{-5}$ and a fixed $\alpha$ of 0.1. 
    A consistent observation was noted wherein the Gauss-Output mechanism applied to SLL models consistently outperformed the Gauss-Input mechanism across all privacy levels.
    Additionally, the Gauss-Input mechanism on a fine-tuned ResNet-18 model demonstrated superior performance compared to the same mechanism on a ResNet-18 model without fine-tuning.     
    In the high privacy region (characterized by low $\epsilon$ values), , the Gauss-Output mechanism on SLL model exhibited clear advantages, whereas in the low privacy region (characterized by high $\epsilon$ values), the Gauss-Input mechanism on  fine-tuned ResNet-18 model showed notable benefits.

\begin{figure}
        \centering
        \includegraphics[width = 0.48 \textwidth]{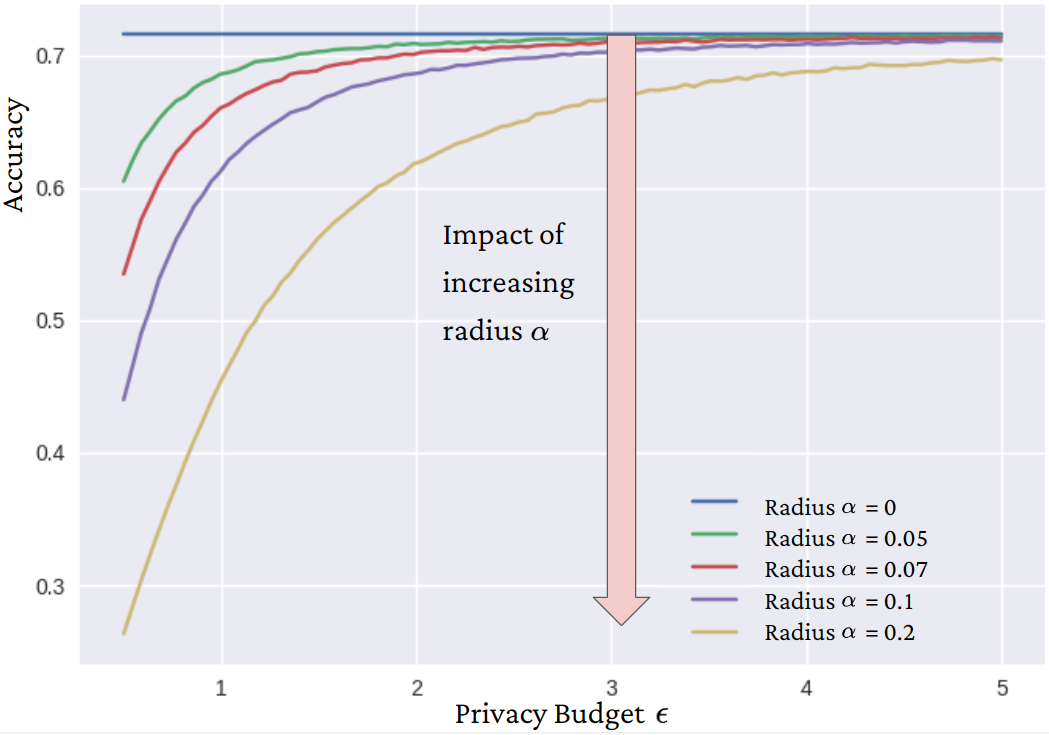}
        \caption{Experimental results on CIFAR-10 classification: natural classification accuracy for SLL models with Gauss-Output mechanism as a function of the privacy budget $\epsilon$ for different values of radius $\alpha$ and a fixed $\delta = 10^{-5}$. Values reported are average of 15 tests. }
        \label{Increasing_radius}
        \vspace{-10pt}
    \end{figure}

\begin{figure*}
        \centering
        \includegraphics[width = \textwidth]{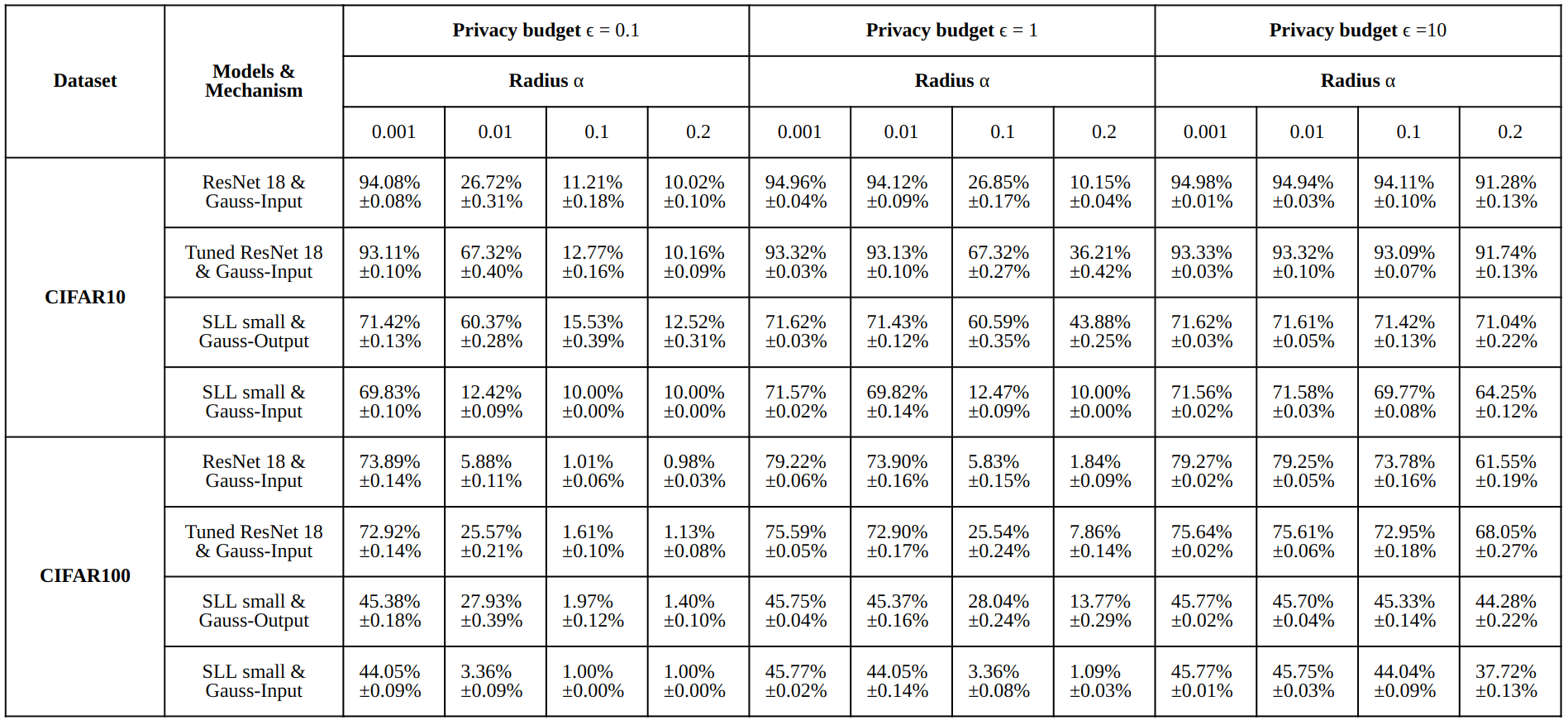}
        \caption{Comparison of the natural accuracy of on CIFAR10 dataset and CIFAR100 dataset under different IP constraints, where $\delta$ is fixed at $10^{-5}$. Values reported are average of 15 tests. }
        \label{table}
        \vspace{-10pt}
    \end{figure*}

    \noindent \textbf{Performance under various IP requirements.} 
    In Figure \ref{Increasing_radius}, we present an analysis of the trade-off between utility and privacy requirements for a fixed mechanism. 
    Specifically, we examine the application of the Gauss-Output mechanism to a SLL model to explore the relationship between model accuracy and radius $\alpha$ as privacy budget $\epsilon$ varies.
    Our observations reveal a three-way trade-off involving utility, radius $\alpha$, and privacy budget $\epsilon$: an increase in the radius $\alpha$ increases, or a decrease in the privacy budget $\epsilon$ leads to a reduction in model utility. 
    As presented in Table \ref{table}, it is observed that the classification accuracy associated with a fixed parameter of $\alpha = 0.1$ is closely aligned with the classification accuracy obtained when $\alpha =0.01$. 
    Notably, the accuracy for $\alpha = 0.01$ and $\epsilon = 0.1$ precisely corresponds to the accuracy for $\alpha = 0.1$ and $\epsilon = 1$.
    In general, since we are injecting Gaussian noise drawn from $\mathcal{N}(0, \sigma^2)$, where $\sigma^2 = 2\ln\left(\frac{1.25}{\delta}\right)\frac{(\alpha \mu_C)^2}{\epsilon^2}$, the noise is directly determined by the ratio $\frac{\alpha}{\epsilon}$. 
    This relationship justifies the observed chaining property as well as the trade-off between privacy radius and privacy budget. 
    Specifically, for a given level of utility, an increasing in the privacy budget requires a corresponding reduction in the privacy radius.
    
\section{Discussion and Future Work}

    \noindent In this paper, we introduced Inference Privacy, a new framework aimed at safeguarding user privacy during the inference phase. 
    We demonstrated its basic properties, supported by theoretical guarantees and empirical results. Several potential avenues for future research are outlined below:

    \noindent a) Utilizing Local Lipschitz constants instead of Global Lipschitz constants for designing IP mechanisms can potentially improve the trade-off between privacy and utility, given that Local Lipschitz constants are typically smaller.
    
    \noindent b) Image classification represents just one application domain of the IP framework. 
    Unlike other privacy frameworks, IP offers flexibility in defining the metric $d$ and privacy radius  $\alpha$ across various contexts, making it promising for privacy preservation in language models.
    
    \noindent c) Leveraging the post-processing characteristic of the proposed IP framework, incorporating noise into the initial segment of a hybrid model \cite{zhong2024splitz}—where only the first half maintains a 1-Lipschitz property—may yield enhanced trade-offs between privacy and utility.

\section*{Acknowledgement}
    \noindent This work was supported by NSF grants CCF 2100013, CNS 2209951, CCF 1651492, CNS 2317192, CNS 1822071, by the U.S. Department of Energy, Office of Science, Office of Advanced Scientific Computing under Award Number DE-SC-ERKJ422, and by NIH through Award 1R01CA261457-01A1.

\bibliographystyle{IEEEtran} 
\bibliography{Reference}

\appendix
\section{Appendix}

\subsection{Proof of Proposition 1}
\label{Post Processing}
    \noindent In this section, we prove the post processing property of IP mechanisms. 
    
    \noindent For arbitrary functions $M : \mathbb {R}^n \rightarrow \mathbb{R}^k $ and $F : \mathbb {R}^k \rightarrow \mathbb{R}^{k'} $, if $M$ satisfies $\{(\epsilon, \delta), \alpha \}$ IP, we need to show the composition $F\circ M$ also satisfies $\{(\epsilon, \delta), \alpha \}$ IP. 
    
    \noindent Consider two arbitrary inputs $x_a, x_b$ such that $d(x_a, x_b) \leq \alpha$, and any measurable subsets $T$ of the output space of $F \circ M$, we need to show that:
    \begin{align}
        Pr[F\circ M(x_a) \in T] \leq e^{\epsilon} Pr[F \circ M(x_b) \in T] + \delta. 
    \end{align}
    
    \noindent Consider the definition of $F\circ M$:
    \begin{align}
        Pr[F \circ M(x_a) \in T] =Pr[F ( M(x_a)) \in T] .
    \end{align}
    Let $S = \{x: F(x) \in T \}$ be the preimage of $T$ under $F$, and $S$ is a measurable subset of the output space of $M$. 
    Therefore:
    \begin{align}
        Pr[F ( M(x_a) ) \in T] =Pr[M(x_a) \in S].
    \end{align}
    Substituting: 
    \begin{align}
        Pr[F ( M(x_a) ) \in T] = & \Pr[M(x_a) \in S] \nonumber \\
        \leq & \ e^{\epsilon} Pr[M(x_b) \in S] + \delta \nonumber \\
        = & \ e^{\epsilon} Pr [F(M(x_b) ) \in T] + \delta .
    \end{align}
    This proves that $F\circ M : \mathbb {R}^n \rightarrow \mathbb{R}^{k'} $ also satisfies $\{(\epsilon, \delta), \alpha \}$ IP, demonstrating the post processing property of IP. 

\subsection{Proof of Proposition 2}
\label{Basic Composition}
    \noindent In this section, we prove the basic composition property of independent $\{(\epsilon, \delta), \alpha \}$ IP mechanisms in an Euclidean $p$-space. 
    
    \noindent Let $M_i : \mathbb {R}^n \rightarrow \mathbb{R}^k $ be an $\{(\epsilon_i, \delta_i), \alpha_i\}$ IP algorithm, and $M$ is defined to be $M(x) = (M_1(x), ..., M_m(x))$, where each $M_i$ are independent from each other. 
    We need to show that $M$ satisfies $\{ ( \sum_{i=1}^m \epsilon_i,  \sum_{i=1}^m \delta_i, \min \alpha_i \}$ IP. 
    
    \noindent Consider two arbitrary inputs $x_a, x_b$ such that  $ d({x}_a, {x}_b) \leq \min \alpha_i $, and any measurable subsets $S$ of the output space of $M$, we need to show:
    \begin{align}
        Pr[M({x}_a)\in S] \leq & e^{\sum^{m}_{i=1} \epsilon_i} Pr[M({x}_b) \in S] + \sum^{m}_{i=1} \delta_i  .
    \end{align}

    \noindent Since $M(x) = (M_1(x), ..., M_m(x))$, the output of $M$ consists all $m$ outputs of $M_1(x), ..., M_m(x)$. 
    We then write $S$ as a subset of the product space of the outputs of $M_1(x), ..., M_m(x)$, that is $S \subseteq  \prod_{i = 1}^{m} S_i$, for any measurable subset $S_i$ of the output space of $M_i$. 

    \noindent Moreover, since for all $\alpha_i \geq \min \alpha_i$, it follows that:
    \begin{align}
        B(x, \min \alpha_i ) \subseteq B(x,\alpha_i) . 
    \end{align}
    Thus, since $M_i$ satisfies $\{(\epsilon, \delta), \alpha_i \}$ IP, it implies that all $M_i$ also satisfies $\{(\epsilon, \delta), \min \alpha_i \}$ IP. 
    Since each mechanism $M_i$ is independent, then:
    \begin{align}
        & \ Pr[M(x_a) \in S] \nonumber \\
        = & \ Pr[ (M_1({x}_a), ..., M_m({x}_a)) \in (S_1, ..., S_m)]  \nonumber \\
        = & \ \prod_{i = 1}^{m} Pr[ (M_i({x}_a) \in S_i] \nonumber \\
        \leq & \ \prod_{i = 1}^{m} \min \{ e^{\epsilon_i} Pr[ (M_i({x}_b) \in S_i] +\delta_i, 1\} \nonumber \\
        \leq & \ \prod_{i = 1}^{m} e^{\epsilon_i} \min \{Pr[ (M_i({x}_b) \in S_i] +\delta_i, 1\} \nonumber \\
        \leq & \ e^{\sum^{m}_{i=1} \epsilon_i} Pr[ (M_1({x}_b), ..., M_m({x}_b)) \in S] + \sum^{m}_{i=1} \delta_i, 
    \end{align}
    for all $x_a, x_b$ such that $d(x_a, x_b) \leq \min \alpha_i$. 
    This proves that mechanism $M$ satisfies $\{ ( \sum_{i=1}^m \epsilon_i,  \sum_{i=1}^m \delta_i, \min \alpha_i \}$ IP, demonstrating the basic composition of independent IP Mechanisms. 

\subsection{Proof of Proposition 3}
\label{Parallel Composition}
    
    \noindent In this section, we prove the parallel composition property of independent $\{(\epsilon, \delta), \alpha \}$ IP mechanisms in an Euclidean $p$-space. 
    
    \noindent Let an arbitrary input $x \in \mathbb{R}^n$ be spited into $m$ disjoint chunks such that $x = x_1 \cup x_2 ... \cup x_m $ and $x_i \cap x_j = \emptyset$ for any $x_i \neq x_j$ , where each $x_i \in \mathbb{R}^{n_i}$ and $\sum^m_{i=1} n_i = n$. 
    Let $M_i : \mathbb{R}^{n_i} \rightarrow \mathbb{R}^{k_i}$ be an $\{(\epsilon_i, \delta_i), \alpha_i\}$ IP algorithm for each partition $x_i \in \mathbb{R}^{n_i}$, and $M$ is defined to be $M(x) = (M_1(x_1), ..., M_m(x_m))$, where each $M_i$ are independent from each other. 
    We need to show $M$ satisfies $\{ ( \sum_{i=1}^m \epsilon_i,  \sum_{i=1}^m \delta_i), \min \alpha_i \}$ IP. 

    \noindent Since $M(x) = (M_1(x), ..., M_m(x))$, the output of $M$ consists all $m$ outputs of $M_1(x), ..., M_m(x)$. 
    We then write $S$ as a subset of the product space of the outputs of $M_1(x), ..., M_m(x)$, that is $S \subseteq  \prod_{i = 1}^{m} S_i$, for any measurable subset $S_i$ of the output space of $M_i$. 
    
    \noindent Consider the input $x_a$, $x_b$ and the corresponding $l_p$ distance $d(x_a, x_b) = {||x_a - x_b||}_p$:
    \begin{align}
       & \ ||x_a - x_b||_p = (\sum^n_{i=1} {|x_a (i) - x_b(i)|}^p)^{1/p} \nonumber \\
       = & \ (\sum^{n_1}_{i=1} {|x_a (i) - x_b(i)|}^p + \sum^{n_1 + n_2}_{i= n_1+ 1} {|x_a (i) - x_b(i)|}^p +... \nonumber \\
       & + \sum^{n_1+...n_m}_{i=n_1+...n_{m-1} + 1} {|x_a (i) - x_b(i)|}^p )^{1/p} \nonumber \\
       = & \ (||{x_a}_1 - {x_b}_1||^p _p + ... + ||{x_a}_m - {x_b}_m||^p _p)^{1/p} \nonumber \\
       = & \ (\sum^m_{i=1} \alpha_i^p)^{1/p} .
    \end{align} 
    Since each $\alpha_i $ are non-negative, then: 
    \begin{align}
        d(x_a, x_b) = (\sum^m_{i=1} \alpha_i^p)^{1/p} \geq \alpha_i \geq \min \alpha_i
    \end{align}
    Moreover, it follows that:
    \begin{align}
        B(x_i, \min \alpha_i ) \subseteq B(x_i,\alpha_i)
    \end{align}
    Thus, since $M_i$ satisfies $\{(\epsilon, \delta), \alpha_i \}$ IP, it implies that all $M_i$ also satisfies $\{(\epsilon, \delta), \min \alpha_i \}$ IP. Since each $x_i$ are disjoint and independent and each $M_i$ are independent, then:
    \begin{align}
        & \ Pr[M(x_a) \in S] \nonumber \\
        = & \ Pr[ (M_1({x_1}_a),  ... , M_m({x_m}_a)) \in (S_1,  ... , S_m)]  \nonumber \\
        = & \ \prod_{i = 1}^{m} Pr[ (M_i({x_i}_a) \in S_i] \nonumber \\
        \leq & \ \prod_{i = 1}^{m} \min \{ e^{\epsilon_i} Pr[ (M_i({x_i}_b) \in S_i] +\delta_i, 1\} \nonumber \\
        \leq & \ \prod_{i = 1}^{m} e^{\epsilon_i} \min \{Pr[ (M_i({x_i}_b) \in S_i] +\delta_i, 1\} \nonumber \\
        \leq & \ e^{\sum^{m}_{i=1} \epsilon_i} Pr[ (M_1({x_1}_b), ..., M_m({x_m}_b)) \in S] + \sum^{m}_{i=1} \delta_i, 
    \end{align}
    for all ${x_a}$, ${x_b}$ such that $d(x_a, x_b) \leq \min \alpha_i$. 
    This proves that mechanism $M$ satisfies $\{ ( \sum_{i=1}^m \epsilon_i,  \sum_{i=1}^m \delta_i, \min \alpha_i \}$ IP, demonstrating the parallel composition of independent IP Mechanisms. 

\subsection{Proof of Proposition 4}
\label{Chaining Property}
    \noindent In this section, we prove the chaining property of $\{(\epsilon, \delta), \alpha \}$ IP mechanism in a Euclidean $p$-space. 
    Let $M : \mathbb{R}^{n} \rightarrow \mathbb{R}^{k}$ be an $\{(\epsilon, \delta), \alpha\}$ inference private algorithm. 
    We need to show that $M$ satisfies $\{(\ceil{\frac{\beta}{\alpha}}\epsilon,  [\frac{e^{\ceil{\frac{\beta}{\alpha}}\epsilon}-1}{e^{\epsilon}-1} ] \delta), \beta \}$ IP for any $\beta \geq 0$.
    
    \noindent Consider two points $x_a, x_b \in \mathbb{R}^n$ such that $||x_a - x_b||_p = \beta \leq \alpha$. 
    Since $M$ satisfies $\{(\epsilon, \delta), \alpha \}$ inference privacy, by definition, when $\beta \leq \alpha$, $\ceil{\frac{\beta}{\alpha}}=1$, $\{(\epsilon, \delta), \alpha\}$ IP implies $\{\epsilon, \delta), \beta \}$ IP. 
    
    \begin{figure}
        \centering
        \includegraphics[width= 0.35 \textwidth]{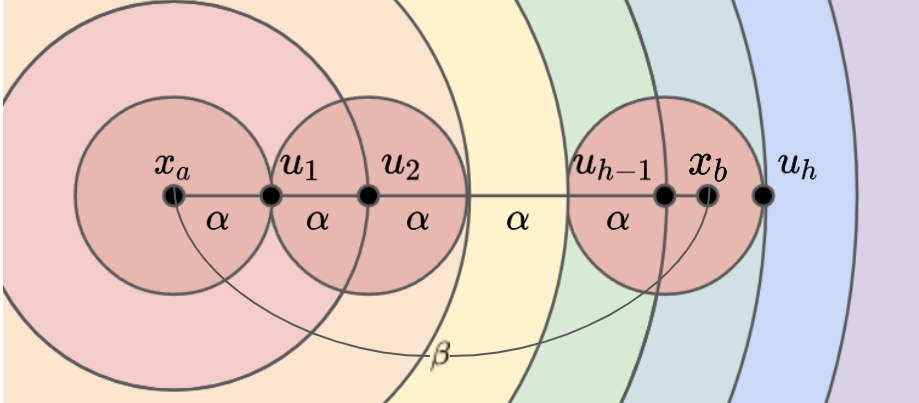}
        \caption{An illustration of 2 points $x_a, x_b$ in an Euclidean $p$-space with distance $\beta > \alpha$}
        \label{Chaining figure}
    \end{figure}
    
    \noindent Illustrated as Figure \ref{Chaining figure}, we consider two points $x_a, x_b \in \mathbb{R}^n$ such that $||x_a - x_b||_p = \beta > \alpha$, and we denote $\ceil{\frac{\beta}{\alpha}} = h \in \mathbb{Z}$, thus:
    \begin{align}
        & (h-1) \alpha < \beta \leq h \alpha .
    \end{align}
    \noindent By definition, the Euclidean $p$-space is a \textit{geodesics space} such that any component of a shortest-length curve between two points lies completely in \cite{busemann2012geometry}. 
    Since all points on the linear segment between $x_a$ and $x_b$ lie in this space, consider a point $u_1$ on this linear segment and that $||x_a-u_1||_p = \alpha$. 
    By definition we have:
    \begin{align}
        ||x_a-u_1||_p + ||u_1 - x_b||_p = & \ ||x_a - x_b||_p \nonumber \\
        ||u_1- x_b ||_p = & \ \beta - \alpha .
    \end{align} 
    Apparently $d(x_a, u_1) \leq \alpha$, for $M$ satisfies $\{(\epsilon, \delta), \alpha \}$ IP:
    \begin{align}
        Pr[M(x_a)\in S] \leq & \ e^{\epsilon} { Pr[M(u_1)\in S]}+ \delta .
    \end{align}
    Then we consider a point $u_2$ on this linear segment and that $||x_a-u_2||_p = 2 \alpha$. 
    We have:
    \begin{align}
        ||x_a-u_1||_p + ||u_1 - u_2||_p = & \ ||x_a - u_2||_p \nonumber \\
        ||u_1- u_2 ||_p = & \ 2\alpha - \alpha = \alpha .
    \end{align} 
    Apparently $d(u_1, u_2) \leq \alpha$, for $M$ satisfies $\{(\epsilon, \delta), \alpha \}$ IP. 
    \noindent In general, we consider $(h-1)$ points $u_1, u_2, ... u_{h-1}$ on this linear segment $s.t.$ $||x_a-u_i||_p = i \alpha$:
    \begin{align}
        ||x_a-u_{i-1}||_p + ||u_{i-1} - u_{i}||_p = & \ ||x_a - u_{i}||_p \nonumber \\
        ||u_{i-1}- u_i ||_p = & \ \alpha .
    \end{align} 
    Apparently $d(u_i, u_{i-1}) \leq \alpha$, for $M$ satisfies $\{(\epsilon, \delta), \alpha \}$ IP.
    Notice that $\beta \leq h \alpha$:
    \begin{align}
        ||x_a-u_{h-1}||_p + ||u_{h-1} -x_b||_p = & \ ||x_a - x_b||_p \nonumber \\
        ||u_{h-1}- x_b ||_p = & \ \beta - (h-1)\alpha \leq \alpha .
    \end{align} 
    Thus $d(x_b, u_{h-1}) \leq \alpha$, for $M$ satisfies $\{(\epsilon, \delta), \alpha \}$ IP.
    In general, we have shown that there exist a sequence of points $u_1, u_2, ... u_{h-1}$ on the linear segment between $x_a$ and $x_b$ such that:
    \begin{align}
        Pr[M(x_a)\in S] \leq & \ e^{\epsilon} { Pr[M(u_1)\in S]}+ \delta \nonumber \\
        Pr[M(u_1)\in S] \leq & \ e^{\epsilon} { Pr[M(u_2)\in S]}+ \delta \nonumber \\
        ...... \nonumber \\
        Pr[M(u_{h-2})\in S] \leq & \ e^{\epsilon} { Pr[M(u_{h-1})\in S]}+ \delta \nonumber \\
        Pr[M(u_{h-1})\in S] \leq & \ e^{\epsilon} { Pr[M(x_b)\in S]}+ \delta \nonumber .
    \end{align} 
    By substitution:
    \begin{align}
        Pr[M(x_a)\in S] \leq & \ e^{\epsilon} ({e^{\epsilon} { Pr[M(u_2)\in S]}+ \delta})+ \delta \nonumber \\
        =  & \ e^{ 2\epsilon} { Pr[M(u_2)\in S]}+ \frac{e^{2 \epsilon}-1}{e^\epsilon -1}\delta \nonumber \\
        \leq & \ {e^{ 2\epsilon}({e^{\epsilon} { Pr[M(u_3)\in S]}+ \delta})+ \delta} + e^{\epsilon} \delta \nonumber \\
        =  & \ e^{ 3\epsilon} { Pr[M(u_3)\in S]}+ \frac{e^{ 3 \epsilon}-1}{e^\epsilon -1}\delta \nonumber \\
        ......
    \end{align}
    For any $\{u_i\}^{h-1}_{i = 1}$, we have:
    \begin{align}
        & \ Pr[M(x_a)\in S] \nonumber \\
        \leq & \ {e^{i \epsilon} { Pr[M(u_i)\in S]}+ \frac{e^{i \epsilon}-1}{e^\epsilon -1}\delta} \nonumber \\
        \leq & \ {e^{i \epsilon} ( {e^{ \epsilon} Pr[M(u_{i+1})\in S] + \delta})+ \frac{e^{i \epsilon}-1}{e^\epsilon -1}\delta} \nonumber \\
        = & \ e^{(i+1) \epsilon} Pr[M(u_{i+1})\in S] + \frac{e^{ (i+1) \epsilon}-1}{e^\epsilon -1}\delta .
    \end{align}
    For $u_{h-1}$, we have:
    \begin{align}
        & \ Pr[M(x_a)\in S] \nonumber \\
        \leq & \ {e^{(h-1) \epsilon} { Pr[M(u_{h-1})\in S]}+ \frac{e^{{(h-1)} \epsilon}-1}{e^\epsilon -1}\delta} \nonumber \\
        \leq & \ {e^{{(h-1)} \epsilon} ( {e^{ \epsilon} Pr[M(x_b)\in S] + \delta})+ \frac{e^{(h-1) \epsilon}-1}{e^\epsilon -1}\delta} \nonumber \\
        = & \ e^{h \epsilon} Pr[M(x_b)\in S] + \frac{e^{ h \epsilon}-1}{e^\epsilon -1}\delta .
    \end{align}

    \noindent This proves that $M$ also satisfies  $\{(\ceil{\frac{\beta}{\alpha}} \epsilon, [\frac{e^{\ceil{\frac{\beta}{\alpha}} \epsilon}-1} {e^{\epsilon}-1} ]\delta), \beta \}$ IP, demonstrating the chaining property of IP mechanisms. 
    
\subsection{Proof of Theorem 1}
\label{Lap-Output}

    \noindent Given any arbitrary pre-trained function $C(\cdot)$ that takes input $x$ and outputs $C(x) \in \mathbb{R}^k $, the Lap-Output Mechanism is defined as:
        \vspace{-10pt}
        \begin{align}
            M_{Lo}(x) = C(x) + Z_1 ,
        \end{align}
        where $Z_1$ is a $k$-dimensional vector with i.i.d. entries drawn from $ Lap(\frac{\mu_C \alpha}{\epsilon})$ and $\mu_C$ is the $\ell_1$ global Lipschitz constant. 
    \noindent We want to show that the Lap-Output Mechanism satisfies $\{ (\epsilon, 0), \alpha \}$ IP. 

    \begin{proof}
        
        Let $x_a \in \mathbb{R}^d$ and $x_b \in \mathbb{R}^d$ be two different arbitrary points such that $d(x_a, x_b) \leq \alpha$. 
        To prove the Lap-Output mechanism satisfies $\{ (\epsilon, 0), \alpha \}$ inference privacy, we need to show that for any output $y \in \mathbb {R}^k$ of the output space of $M_{Lo}$: 
        \begin{align}
             Pr[M_{Lo}(x_a) = y] \leq & \ e^{\epsilon} { Pr[M_{Lo}(x_b) = y]} \nonumber .
        \end{align}
        Which is equivalent to:
        \begin{align}
            \frac{Pr[M_{Lo}(x_a) = y]} {Pr[M_{Lo}(x_b) =y]} \leq & \ e^{\epsilon} \nonumber .
        \end{align}
        We look at this ratio, by definition of $M_{Lo}$:
        \begin{align}
            \frac{Pr[M_{Lo}(x_a) = y]} {Pr[M_{Lo}(x_b) = y]} = & \ \frac{Pr[C(x_a)+Z_1 = y]} {Pr[C(x_b) +Z_1 =y ]}  \nonumber\\
            = & \ \frac{Pr[Z_1 = y - C(x_a) ]} {Pr[Z_1 = y - C(x_b)]} . 
        \end{align}
        Since $Z_1 \in \mathbb {R}^k$, we denote $Z_1 = (z_1, z_2, ..., z_k)$, where $z_i$ represents the i-th entry of $Z_1$. 
        Then:
        \begin{align}
            Pr[Z_1 = z] = \prod_{i=1}^k \frac{\epsilon}{2 \mu_{C} \alpha} \exp \left( {-\frac{\epsilon |z_i|}{\alpha \mu_C}} \right). 
        \end{align}
        Then:
        \begin{align}
            & \ \frac{Pr[M_{Lo}(x_a)\in S]} {Pr[M_{Lo}(x_b)\in S]} \\
            =  & \ \prod_{i=1}^k \left( \frac{\exp ({-\frac{\epsilon |C(x_a)_i-z_i|}{\alpha \mu_C}})}{\exp (- \frac{\epsilon |C(x_b)_i-z_i|}{\alpha \mu_C})}\right)  \nonumber\\
            = & \ \prod_{i=1}^k \exp \left( \frac{\epsilon (|C(x_b)_i - z_i|- |C(x_a)_i - z_i| )}{\alpha \mu_C}\right)  \nonumber\\
            \leq & \ \prod_{i=1}^k \exp \left( \frac{\epsilon |C(x_a)_i - C(x_b)_i|}{\alpha \mu_C}\right) \nonumber\\
            \leq & \ \exp(\epsilon) .
        \end{align}
        Where the first inequality follows from triangle inequality:
        \begin{align}
            |C(x_b)_i - z_i|- |C(x_a)_i - z_i| \leq |C(x_a)_i - C(x_b)_i|. 
        \end{align}
        And the last inequality follows from the fact that:
        \begin{align}
            \alpha \mu_C  \geq ||\Delta_f (x_a, x_b)||_1 .
        \end{align}
        Which proves that Lap-Output Mechanism satisfies $\{ (\epsilon, 0), \alpha \}$ IP. 
    
    \end{proof}

\subsection{Proof of Theorem 2}
\label{Gauss-Output}

    \noindent Given any function $C(\cdot)$ takes input $x \in \mathbb{R}^n$ and outputs $C(x) \in \mathbb{R}^k $, the Gauss-Output Mechanism is defined as: 
        \begin{align}
            M_{Go}(x) = C(x) + Z_2, 
        \end{align}
    where $Z_2$ is a $k$-dimensional vector with i.i.d. entries drawn from $ \mathcal{N}(0, \sigma^2)$ ,  $\sigma^2 = 2ln(\frac{1.25}{\delta})\frac{(\alpha {\mu_C)} ^2}{\epsilon^2 }$ and $\mu_C$ is the $l_2$  global Lipschitz constant of $C(x)$.
    \noindent We want to show the Gauss-Output Mechanism satisfies $\{ (\epsilon, \delta), \alpha \}$ IP. 
   \begin{proof}
   
        Let $x_a \in \mathbb{R}^d $ and $x_b \in \mathbb{R}^d$ be two different arbitrary points such that $d (x_a, x_b) \leq \alpha $.
        To prove the Gauss-Output mechanism satisfies $\{ (\epsilon, \delta), \alpha \}$ inference privacy, we need to show that for any measurable subset $S$ of the output space of $M_{Go}$:
        \begin{align}
            Pr[M_{Go}(x_a)\in S] \leq & \ e^{\epsilon} { Pr[M_{Go}(x_b)\in S]}+ \delta
        \end{align}
        Now we define $S' = \{s - C(x_a) : s \in S \}$, then each probability can be expressed as an integration:
        \begin{align}
            I_1 = & \ Pr[M_{Go}(x_a)\in S] = Pr[Z_2\in S'] \nonumber \\
            = & \ \int\limits_{Z_2\in S'} f_z(Z_2) dz .
        \end{align}
        And:
        \begin{align}
            I_2 = & \ Pr[M_{Go}(x_b)\in S] = Pr[C(x_b) - C(x_a) + Z_2\in S'] \nonumber \\
            = & \ \int\limits_{ C(x_b) - C(x_a) + Z_2\in S'} f_z(Z_2) du \nonumber \\
            = & \ \int\limits_{Z_2 \in S'} f_z(Z_2-(C(x_b) - C(x_a)) dz .
        \end{align}
        We then partition the entire $\mathbb{R}^k$ into two parts as $\mathbb{R}_1^k \cup \mathbb{R}_2^k $, where:
        
        \begin{align}
            \mathbb{R}_1^k = \{ x\in \mathbb{R}^k :|ln \left( \frac{f_z(Z_2)}{f_z(Z_2 - (C(x_b) - C(x_a)))} \right)| \leq \epsilon\} \nonumber \\
            \mathbb{R}_1^k = \{ x\in \mathbb{R}^k :|ln \left( \frac{f_z(Z_2)}{f_z(Z_2 - (C(x_b) - C(x_a)))} \right)| > \epsilon\} .
        \end{align}
        For any fixed subset $S' \in \mathbb{R}^k$, define$S' = S'_1 \cup S'_2$ where:
        \begin{align}
            S'_1= {Z_2 \in \mathbb{R}_1^k} \nonumber \\
            S'_2= {Z_2 \in \mathbb{R}_2^k} .
        \end{align}
        Then: 
        \begin{align}
            I_1 = & \ \int\limits_{Z_2\in S_1'} f_z(Z_2) dz + \int\limits_{Z_2\in S_2'} f_z(Z_2) dz \nonumber\\
            \leq & \ e^{\epsilon}  \int\limits_{Z_2\in S_1'} f_z(Z_2 - (C(x_a) -C(x_b))) dz + \int\limits_{Z_2\in S_2'} f_u(Z_2) dz .
        \end{align}
        Notice that:
        \begin{align}
            & \int\limits_{Z_2\in S_2'} f_z(Z_2) dz \nonumber \\
            = & \int\limits_{Z_2 | ln \left( \frac{f_z(Z_2)}{f_z(Z_2 - (C(x_b) - C(x_a))} \right) > \epsilon} f_u(Z_2) dz .
        \end{align}
        We then look at the interested ratio: 
        \begin{align}
            & \ ln \left( \frac{f_z(Z_2)}{f_z(Z_2 - (C(x_b) - C(x_a)))} \right) \nonumber \\
            = & \ ln \left( \frac{\exp({\sum_{i = 1}^{k}\frac{-z_i^2}{2\sigma^2})}}{\exp ({\sum_{i = 1}^{k}\frac{-(z_i+C(x_b)_i - C(x_a)_i)^2}{2\sigma^2}})} \right) \nonumber \\
            = & \ (-\frac{1}{2\sigma^2})(||z||_2^2 -||Z_2 + C(x_a) - C(x_b)||_2^2) \nonumber\\
            = & \ \frac{||C(x_a) - C(x_b)||_2^2}{2\sigma^2} + \frac{ 2 Z_2 ^T (C(x_a) - C(x_b))}{2 \sigma^2} .
        \end{align}
        Where $ \frac{||C(x_a) - C(x_b)||_2^2}{2\sigma^2}$ is a constant term irreverent of the noise and the term $\ Z_2 ^T (C(x_a) - C(x_b)) $ is a random variable follows $Gaussian (0, \sigma^2 ||C(x_a) - C(x_b)||_2^2 )$. \\
        Letting $Z \sim Gaussian(0, 1)$, the privacy budget can be rewritten as:
        \begin{align}
            \frac{||C(x_a) - C(x_b)||_2}{\sigma} Z + \frac{||C(x_a) - C(x_b)||_2^2}{2\sigma^2} . 
        \end{align}
        Thus, the probability this budget exceeds $\epsilon$ is:
        \begin{align}
            Pr[|Z| > \frac{\sigma \epsilon}{|C(x_a) - C(x_b)||_2} - \frac{||C(x_a) - C(x_b)||_2}{2\sigma}] .
        \end{align}
        Notice that:
        \begin{align}
            \alpha \mu_C \geq ||C(x_a) - C(x_b)||_2 .
        \end{align}
        Thus:
        \begin{align}
            Pr[|Z| > \frac{\sigma \epsilon}{\alpha \mu_C} - \frac{\alpha \mu_C}{2\sigma} ] .
        \end{align}
        Follow a standard Gaussian tail bound, we then have:
        \begin{align}
            \frac{\delta}{2} > \frac{\sigma}{\sqrt{2 \pi} z } e^{-z^2/2\sigma^2} \geq Pr[Z >z]. 
        \end{align}
        To ensure $M_{Go}$ satisfies $\{(\epsilon, \delta), \alpha \}$ inference privacy, we need to set $\sigma \geq \sqrt{2ln(\frac{1.25}{\delta})}\frac{\alpha \mu_C}{\epsilon}$. 
        \begin{align}
            Pr[M_{Go}(x_a)\in S] = & \ { Pr[C(x_a) + Z_2 \in S]} \nonumber\\
            = & \ { Pr[Z_2 \in S'_1]} +  { Pr[Z_2 \in S'_2]} \nonumber\\
            \leq & \ { Pr[Z_2 \in S'_1]} +  \delta \nonumber\\
            \leq & \ e^{\epsilon}  { Pr[Z_2 + C(x_b) - C(x_a) \in S'_1]} +\delta \nonumber \\
            = & \ e^{\epsilon}  { Pr[M_{Go} (x_b) \in S]} +\delta .
        \end{align}
        Then we show $M_{Go}$ also satisfies $\{(\epsilon, \delta), \alpha \}$ inference privacy.
            
    \end{proof}

\subsection{Proof of Theorem 3}
\label{Gauss-Input}

    \noindent Given any function $C(\cdot)$ takes input $x \in \mathbb{R}^n$ and outputs $C(x) \in \mathbb{R}^k $, the Gauss-Input Mechanism is defined as:
        \begin{align}
            M (x) = C(x + Z_2),
        \end{align}
        where $Z_3$ is a $n$-dimensional vector with i.i.d. entries drawn from $ \mathcal{N}(0, \sigma^2)$, and $\sigma^2 = 2ln(\frac{1.25}{\delta})\frac{\alpha^2}{\epsilon^2 }$. 
        We want to show the Gauss-Input mechanism satisfies $\{(\epsilon, \delta), \alpha \}$ IP. 
    \begin{proof}
    
        Let $x_a \in \mathbb{R}^d $ and $x_b \in \mathbb{R}^d$ be two different arbitrary points, such that $d(x_a, x_b) \leq \alpha$. 
        Let $A(\cdot)$ be some arbitrary function $A : \mathbb{R}^d \rightarrow \mathbb{R}^d$ that outputs the models inputs directly:
        \begin{align}
            A(x) = x .
        \end{align}
        Then:
        \begin{align}
            M_{Gi}(x) = C\circ (A(x) + Z_2) .
        \end{align}
        If $A(x) + Z_3 $ preserves $\{ (\epsilon, \delta), \alpha \}$ inference privacy, then $M_{Gi}(x)$ must preserves $\{ (\epsilon, \delta), \alpha \}$ inference privacy.
        We want to show that for any measurable subset $S$ of the output space of $A$:
        \begin{align}
            Pr[A(x_a)\in S] \leq & \ e^{\epsilon} { Pr[A(x_b)\in S]}+ \delta . 
        \end{align}
        Now we define $S' = \{s - x_a : s \in S \}$, then each probability can be expressed as an integration:
        \begin{align}
            I_1 = & \ Pr[A(x_a)\in S] = Pr[Z_3\in S'] \nonumber \\
            = & \ \int\limits_{Z_3\in S'} f_z(Z_3) dz .
        \end{align}
        And:
        \begin{align}
            I_2 = & \ Pr[A(x_b)\in S] = Pr[x_b - x_a + Z_3\in S'] \nonumber \\
            = & \ \int\limits_{x_b - x_a + Z_3\in S'} f_z(Z_3) dz \nonumber \\
            = & \ \int\limits_{Z_3 \in S'} f_z(Z_3-(x_b - x_a) dz .
        \end{align}
        We then partition the entire $\mathbb{R}^d$ into two parts as $\mathbb{R}_1^d \cup \mathbb{R}_2^d $, where:
        \begin{align}
            \mathbb{R}_1^d = \{ x\in \mathbb{R}^d :|ln \left(\frac{f_z(Z_3)}{f_z(Z_3 - (x_b - x_a))}\right)| \leq \epsilon\} \nonumber \\
            \mathbb{R}_1^d = \{ x\in \mathbb{R}^d :|ln \left(\frac{f_z(Z_3)}{f_z(Z_3 - (x_b - x_a))}\right)| > \epsilon\} .
        \end{align}
        For any fixed subset $S' \in \mathbb{R}^d$, define$S' = S'_1 \cup S'_2$ where: 
        \begin{align}
            S'_1= {Z_3 \in \mathbb{R}_1^d} \nonumber \\
            S'_2= {Z_3 \in \mathbb{R}_2^d} .
        \end{align}
        Then: 
        \begin{align}
            I_1 = & \ \int\limits_{Z_3\in S_1'} f_z(Z_3) dz + \int\limits_{Z_3\in S_2'} f_z(Z_3) dz \nonumber\\
            \leq & \ e^{\epsilon}  \int\limits_{Z_3\in S_1'} f_z(Z_3 - (x_a -x_b)) dz + \int\limits_{Z_3\in S_2'} f_z(Z_3) dz .
        \end{align}
        Notice that:
        \begin{align}
            \int\limits_{Z_3\in S_2'} f_z(Z_3) dz = \int\limits_{Z_3 | ln \left( \frac{f_z(Z_3)}{f_z(Z_3 - (x_b - x_a))} \right) > \epsilon} f_z(Z_3) dz .
        \end{align}
        We look at the interested ratio: 
        \begin{align}
            & \ ln \left( \frac{f_z(Z_3)}{f_z(Z_3 - (x_b - x_a))} \right) \nonumber \\
            = & \ ln \left( \frac{\exp({\sum_{i = 1}^{d}\frac{-u_i^2}{2\sigma^2}})}{\exp ({\sum_{i = 1}^{d}\frac{-(z_i+(x_b)_i - (x_a)_i)^2}{2\sigma^2}})} \right) \nonumber \\
            = & \ -\frac{1}{2\sigma^2}(||z||_2^2 -||z + (x_a - x_b)||_2^2) \nonumber\\
            = & \ \frac{||x_a -  x_b||_2^2}{2\sigma^2} + \frac{ 2 Z_3 ^T (x_a -  x_b )}{2 \sigma^2} .
        \end{align}
        Where $ \frac{||x_a - x_b||_2^2}{2\sigma^2}$ is a constant term irreverent of the noise and the term $\ Z_3^T (x_a - x_b) $ is a random variable follows $Gaussian (0, \sigma^2 ||(x_a - x_b )||_2^2 )$. \\
        Letting $Z \sim Gaussian(0, 1)$, the privacy budget can be rewritten as:
        \begin{align}
            \frac{||x_a - x_b||_2}{\sigma} Z + \frac{||x_a - x_b||_2^2}{2\sigma^2} .
        \end{align}
        Thus, the probability this budget exceeds $\epsilon$ is:
        \begin{align}
            Pr[|Z| > \frac{\sigma \epsilon}{||x_a-x_b||_2} - \frac{||x_a-  x_b||_2}{2\sigma}] .
        \end{align}
        Based on definition of $x_a, x_b$, notice that:
        \begin{align}
            \alpha \geq ||x_a - x_b ||_2 .
        \end{align}
        Thus:
        \begin{align}
            Pr[|Z| > \frac{\sigma \epsilon}{\alpha} - \frac{\alpha}{2\sigma} ]
        \end{align}
        Follow a standard Gaussian tail bound we then have :
        \begin{align}
            \frac{\delta}{2} > \frac{\sigma}{\sqrt{2 \pi} z } e^{-z^2/2\sigma^2} \geq Pr[Z >z]. 
        \end{align}
        To ensure $A$ satisfies $\{(\epsilon, \delta), \alpha \}$ IP, we need to set $\sigma \geq \sqrt{2ln(\frac{1.25}{\delta})}\frac{\alpha}{\epsilon}$. 
        \begin{align}
            Pr[A(x_a)\in S] = & \ { Pr[x_a + Z_3 \in S]} \nonumber\\
            = & \ { Pr[Z_3 \in S'_1]} +  { Pr[Z_3 \in S'_2]} \nonumber\\
            \leq & \ { Pr[Z_3 \in S'_1]} +  \delta \nonumber\\
            \leq & \ e^{\epsilon}  { Pr[Z_3 +x_b - x_a \in S'_1]} +\delta \nonumber \\
            = & \ e^{\epsilon}  { Pr[A (x_b) \in S]} +\delta .
        \end{align}
        Then we show $A$ also satisfies $\{(\epsilon, \delta), \alpha \}$ inference privacy, and by post processing property, $M_{Gi}$ also satisfies $\{(\epsilon, \delta), \alpha \}$ inference privacy.

    \end{proof}

\end{document}